\documentclass[sigconf]{acmart}

\usepackage[utf8]{inputenc}
\usepackage{algorithmic}
\usepackage{caption}
\usepackage{color, colortbl}
\usepackage{graphicx}
\usepackage{hyperref}
\usepackage[draft]{minted}
\usepackage{subfig}
\usepackage{textcomp}
\usepackage{xcolor}
\usepackage{booktabs}
\usepackage{enumitem}

\definecolor{LG}{gray}{0.9}
\definecolor{HG}{gray}{0.6}

\newcommand{\administrativeTemplates}{ATs}

\newcommand{\notSecurityRelevant}{\textsc{nsr}}
\newcommand{\securityRelevant}{\textsc{sr}}

\newcommand{\windowsServer}{WS16}
\newcommand{\windowsTen}{W10}

\makeatletter
\def\PYG@reset{\let\PYG@it=\relax \let\PYG@bf=\relax%
    \let\PYG@ul=\relax \let\PYG@tc=\relax%
    \let\PYG@bc=\relax \let\PYG@ff=\relax}
\def\PYG@tok#1{\csname PYG@tok@#1\endcsname}
\def\PYG@toks#1+{\ifx\relax#1\empty\else%
    \PYG@tok{#1}\expandafter\PYG@toks\fi}
\def\PYG@do#1{\PYG@bc{\PYG@tc{\PYG@ul{%
    \PYG@it{\PYG@bf{\PYG@ff{#1}}}}}}}
\def\PYG#1#2{\PYG@reset\PYG@toks#1+\relax+\PYG@do{#2}}

\expandafter\def\csname PYG@tok@gd\endcsname{\def\PYG@tc##1{\textcolor[rgb]{0.63,0.00,0.00}{##1}}}
\expandafter\def\csname PYG@tok@gu\endcsname{\let\PYG@bf=\textbf\def\PYG@tc##1{\textcolor[rgb]{0.50,0.00,0.50}{##1}}}
\expandafter\def\csname PYG@tok@gt\endcsname{\def\PYG@tc##1{\textcolor[rgb]{0.00,0.27,0.87}{##1}}}
\expandafter\def\csname PYG@tok@gs\endcsname{\let\PYG@bf=\textbf}
\expandafter\def\csname PYG@tok@gr\endcsname{\def\PYG@tc##1{\textcolor[rgb]{1.00,0.00,0.00}{##1}}}
\expandafter\def\csname PYG@tok@cm\endcsname{\let\PYG@it=\textit\def\PYG@tc##1{\textcolor[rgb]{0.25,0.50,0.50}{##1}}}
\expandafter\def\csname PYG@tok@vg\endcsname{\def\PYG@tc##1{\textcolor[rgb]{0.10,0.09,0.49}{##1}}}
\expandafter\def\csname PYG@tok@vi\endcsname{\def\PYG@tc##1{\textcolor[rgb]{0.10,0.09,0.49}{##1}}}
\expandafter\def\csname PYG@tok@vm\endcsname{\def\PYG@tc##1{\textcolor[rgb]{0.10,0.09,0.49}{##1}}}
\expandafter\def\csname PYG@tok@mh\endcsname{\def\PYG@tc##1{\textcolor[rgb]{0.40,0.40,0.40}{##1}}}
\expandafter\def\csname PYG@tok@cs\endcsname{\let\PYG@it=\textit\def\PYG@tc##1{\textcolor[rgb]{0.25,0.50,0.50}{##1}}}
\expandafter\def\csname PYG@tok@ge\endcsname{\let\PYG@it=\textit}
\expandafter\def\csname PYG@tok@vc\endcsname{\def\PYG@tc##1{\textcolor[rgb]{0.10,0.09,0.49}{##1}}}
\expandafter\def\csname PYG@tok@il\endcsname{\def\PYG@tc##1{\textcolor[rgb]{0.40,0.40,0.40}{##1}}}
\expandafter\def\csname PYG@tok@go\endcsname{\def\PYG@tc##1{\textcolor[rgb]{0.53,0.53,0.53}{##1}}}
\expandafter\def\csname PYG@tok@cp\endcsname{\def\PYG@tc##1{\textcolor[rgb]{0.74,0.48,0.00}{##1}}}
\expandafter\def\csname PYG@tok@gi\endcsname{\def\PYG@tc##1{\textcolor[rgb]{0.00,0.63,0.00}{##1}}}
\expandafter\def\csname PYG@tok@gh\endcsname{\let\PYG@bf=\textbf\def\PYG@tc##1{\textcolor[rgb]{0.00,0.00,0.50}{##1}}}
\expandafter\def\csname PYG@tok@ni\endcsname{\let\PYG@bf=\textbf\def\PYG@tc##1{\textcolor[rgb]{0.60,0.60,0.60}{##1}}}
\expandafter\def\csname PYG@tok@nl\endcsname{\def\PYG@tc##1{\textcolor[rgb]{0.63,0.63,0.00}{##1}}}
\expandafter\def\csname PYG@tok@nn\endcsname{\let\PYG@bf=\textbf\def\PYG@tc##1{\textcolor[rgb]{0.00,0.00,1.00}{##1}}}
\expandafter\def\csname PYG@tok@no\endcsname{\def\PYG@tc##1{\textcolor[rgb]{0.53,0.00,0.00}{##1}}}
\expandafter\def\csname PYG@tok@na\endcsname{\def\PYG@tc##1{\textcolor[rgb]{0.49,0.56,0.16}{##1}}}
\expandafter\def\csname PYG@tok@nb\endcsname{\def\PYG@tc##1{\textcolor[rgb]{0.00,0.50,0.00}{##1}}}
\expandafter\def\csname PYG@tok@nc\endcsname{\let\PYG@bf=\textbf\def\PYG@tc##1{\textcolor[rgb]{0.00,0.00,1.00}{##1}}}
\expandafter\def\csname PYG@tok@nd\endcsname{\def\PYG@tc##1{\textcolor[rgb]{0.67,0.13,1.00}{##1}}}
\expandafter\def\csname PYG@tok@ne\endcsname{\let\PYG@bf=\textbf\def\PYG@tc##1{\textcolor[rgb]{0.82,0.25,0.23}{##1}}}
\expandafter\def\csname PYG@tok@nf\endcsname{\def\PYG@tc##1{\textcolor[rgb]{0.00,0.00,1.00}{##1}}}
\expandafter\def\csname PYG@tok@si\endcsname{\let\PYG@bf=\textbf\def\PYG@tc##1{\textcolor[rgb]{0.73,0.40,0.53}{##1}}}
\expandafter\def\csname PYG@tok@s2\endcsname{\def\PYG@tc##1{\textcolor[rgb]{0.73,0.13,0.13}{##1}}}
\expandafter\def\csname PYG@tok@nt\endcsname{\let\PYG@bf=\textbf\def\PYG@tc##1{\textcolor[rgb]{0.00,0.50,0.00}{##1}}}
\expandafter\def\csname PYG@tok@nv\endcsname{\def\PYG@tc##1{\textcolor[rgb]{0.10,0.09,0.49}{##1}}}
\expandafter\def\csname PYG@tok@s1\endcsname{\def\PYG@tc##1{\textcolor[rgb]{0.73,0.13,0.13}{##1}}}
\expandafter\def\csname PYG@tok@dl\endcsname{\def\PYG@tc##1{\textcolor[rgb]{0.73,0.13,0.13}{##1}}}
\expandafter\def\csname PYG@tok@ch\endcsname{\let\PYG@it=\textit\def\PYG@tc##1{\textcolor[rgb]{0.25,0.50,0.50}{##1}}}
\expandafter\def\csname PYG@tok@m\endcsname{\def\PYG@tc##1{\textcolor[rgb]{0.40,0.40,0.40}{##1}}}
\expandafter\def\csname PYG@tok@gp\endcsname{\let\PYG@bf=\textbf\def\PYG@tc##1{\textcolor[rgb]{0.00,0.00,0.50}{##1}}}
\expandafter\def\csname PYG@tok@sh\endcsname{\def\PYG@tc##1{\textcolor[rgb]{0.73,0.13,0.13}{##1}}}
\expandafter\def\csname PYG@tok@ow\endcsname{\let\PYG@bf=\textbf\def\PYG@tc##1{\textcolor[rgb]{0.67,0.13,1.00}{##1}}}
\expandafter\def\csname PYG@tok@sx\endcsname{\def\PYG@tc##1{\textcolor[rgb]{0.00,0.50,0.00}{##1}}}
\expandafter\def\csname PYG@tok@bp\endcsname{\def\PYG@tc##1{\textcolor[rgb]{0.00,0.50,0.00}{##1}}}
\expandafter\def\csname PYG@tok@c1\endcsname{\let\PYG@it=\textit\def\PYG@tc##1{\textcolor[rgb]{0.25,0.50,0.50}{##1}}}
\expandafter\def\csname PYG@tok@fm\endcsname{\def\PYG@tc##1{\textcolor[rgb]{0.00,0.00,1.00}{##1}}}
\expandafter\def\csname PYG@tok@o\endcsname{\def\PYG@tc##1{\textcolor[rgb]{0.40,0.40,0.40}{##1}}}
\expandafter\def\csname PYG@tok@kc\endcsname{\let\PYG@bf=\textbf\def\PYG@tc##1{\textcolor[rgb]{0.00,0.50,0.00}{##1}}}
\expandafter\def\csname PYG@tok@c\endcsname{\let\PYG@it=\textit\def\PYG@tc##1{\textcolor[rgb]{0.25,0.50,0.50}{##1}}}
\expandafter\def\csname PYG@tok@mf\endcsname{\def\PYG@tc##1{\textcolor[rgb]{0.40,0.40,0.40}{##1}}}
\expandafter\def\csname PYG@tok@err\endcsname{\def\PYG@bc##1{\setlength{\fboxsep}{0pt}\fcolorbox[rgb]{1.00,0.00,0.00}{1,1,1}{\strut ##1}}}
\expandafter\def\csname PYG@tok@mb\endcsname{\def\PYG@tc##1{\textcolor[rgb]{0.40,0.40,0.40}{##1}}}
\expandafter\def\csname PYG@tok@ss\endcsname{\def\PYG@tc##1{\textcolor[rgb]{0.10,0.09,0.49}{##1}}}
\expandafter\def\csname PYG@tok@sr\endcsname{\def\PYG@tc##1{\textcolor[rgb]{0.73,0.40,0.53}{##1}}}
\expandafter\def\csname PYG@tok@mo\endcsname{\def\PYG@tc##1{\textcolor[rgb]{0.40,0.40,0.40}{##1}}}
\expandafter\def\csname PYG@tok@kd\endcsname{\let\PYG@bf=\textbf\def\PYG@tc##1{\textcolor[rgb]{0.00,0.50,0.00}{##1}}}
\expandafter\def\csname PYG@tok@mi\endcsname{\def\PYG@tc##1{\textcolor[rgb]{0.40,0.40,0.40}{##1}}}
\expandafter\def\csname PYG@tok@kn\endcsname{\let\PYG@bf=\textbf\def\PYG@tc##1{\textcolor[rgb]{0.00,0.50,0.00}{##1}}}
\expandafter\def\csname PYG@tok@cpf\endcsname{\let\PYG@it=\textit\def\PYG@tc##1{\textcolor[rgb]{0.25,0.50,0.50}{##1}}}
\expandafter\def\csname PYG@tok@kr\endcsname{\let\PYG@bf=\textbf\def\PYG@tc##1{\textcolor[rgb]{0.00,0.50,0.00}{##1}}}
\expandafter\def\csname PYG@tok@s\endcsname{\def\PYG@tc##1{\textcolor[rgb]{0.73,0.13,0.13}{##1}}}
\expandafter\def\csname PYG@tok@kp\endcsname{\def\PYG@tc##1{\textcolor[rgb]{0.00,0.50,0.00}{##1}}}
\expandafter\def\csname PYG@tok@w\endcsname{\def\PYG@tc##1{\textcolor[rgb]{0.73,0.73,0.73}{##1}}}
\expandafter\def\csname PYG@tok@kt\endcsname{\def\PYG@tc##1{\textcolor[rgb]{0.69,0.00,0.25}{##1}}}
\expandafter\def\csname PYG@tok@sc\endcsname{\def\PYG@tc##1{\textcolor[rgb]{0.73,0.13,0.13}{##1}}}
\expandafter\def\csname PYG@tok@sb\endcsname{\def\PYG@tc##1{\textcolor[rgb]{0.73,0.13,0.13}{##1}}}
\expandafter\def\csname PYG@tok@sa\endcsname{\def\PYG@tc##1{\textcolor[rgb]{0.73,0.13,0.13}{##1}}}
\expandafter\def\csname PYG@tok@k\endcsname{\let\PYG@bf=\textbf\def\PYG@tc##1{\textcolor[rgb]{0.00,0.50,0.00}{##1}}}
\expandafter\def\csname PYG@tok@se\endcsname{\let\PYG@bf=\textbf\def\PYG@tc##1{\textcolor[rgb]{0.73,0.40,0.13}{##1}}}
\expandafter\def\csname PYG@tok@sd\endcsname{\let\PYG@it=\textit\def\PYG@tc##1{\textcolor[rgb]{0.73,0.13,0.13}{##1}}}

\def\PYGZbs{\char`\\}
\def\PYGZus{\char`\_}

\def\PYGZhy{\char`\-}

\def\PYGZdq{\char`\"}

\makeatother

\copyrightyear{2022}
\acmYear{2022}
\setcopyright{acmcopyright}
\acmConference[ASE '22]{37th IEEE/ACM International Conference on Automated Software Engineering}{October 10--14, 2022}{Rochester, MI, USA}
\acmBooktitle{37th IEEE/ACM International Conference on Automated Software Engineering (ASE '22), October 10--14, 2022, Rochester, MI, USA}
\acmPrice{15.00}
\acmDOI{10.1145/3551349.3559499}
\acmISBN{978-1-4503-9475-8/22/10}

\begin{document}

\title{Automated Identification of Security-Relevant Configuration Settings Using NLP}

\author{Patrick Stöckle}
\email{patrick.stoeckle@tum.de}
\orcid{0000-0003-0193-5871}

\affiliation{%
  \institution{Technical University of Munich}
  \city{Munich}
  \country{Germany}
}

\author{Theresa Wasserer}
\email{theresa.wasserer@tum.de}
\orcid{0000-0001-9699-0587}
\affiliation{%
  \institution{Technical University of Munich}
  \city{Munich}
  \country{Germany}
}

\author{Bernd Grobauer}
\email{bernd.grobauer@siemens.com}
\orcid{0000-0003-0792-3935}
\affiliation{%
  \institution{Siemens AG}
  \city{Munich}
  \country{Germany}
}

\author{Alexander Pretschner}
\email{alexander.pretschner@tum.de}
\orcid{0000-0002-5573-1201}
\affiliation{%
  \institution{Technical University of Munich}
  \city{Munich}
  \country{Germany}
}

\begin{abstract}
To secure computer infrastructure, we need to configure all security-relevant settings.
We need security experts to identify security-relevant settings, but this process is time-consuming and expensive.
Our proposed solution uses state-of-the-art natural language processing to classify settings as security-relevant based on their description.
Our evaluation shows that our trained classifiers do not perform well enough to replace the human security experts but can help them classify the settings.
By publishing our labeled data sets and the code of our trained model, we want to help security experts analyze configuration settings and enable further research in this area.
\end{abstract}




\maketitle

\section{Introduction}

\begin{figure}
\centerline{\includegraphics[width=0.7\columnwidth]{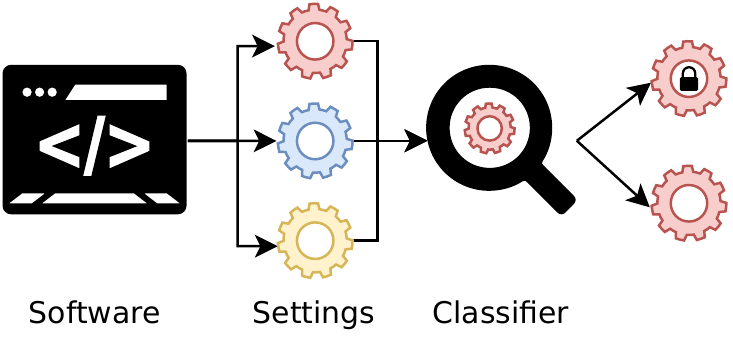}}
\caption{Identification of security-relevant settings.}
\Description{
The graphic goes from left to right.
Left is the software.
Three arrows lead from the software to three configuration settings, one in red, one in yellow, and one in blue.
Three arrows lead from the settings to a magnifier with a setting in focus, symbolizing the classifier.
Two arrows lead from the magnifier to two settings, one with a lock, i.e., a security-relevant setting, and one without a lock, i.e., a non-security-relevant setting.
}
\label{fig:problem}
\end{figure}

A critical part of the IT security in an organization such as Siemens is the secure configuration of all used software~\cite{dietrich2018investigating}.
Here, we need to know which configuration settings (from here on \textit{settings}) of a software are security-relevant (\securityRelevant{}) or not security-relevant (\notSecurityRelevant{}) (see~\autoref{fig:problem}).
We denote the classification predicate with $ p $.
Going through all possible settings $ \Gamma_\theta $ of a software $ \theta $ and classifying whether a setting $ \gamma \in \Gamma_\theta $ is \securityRelevant{} ($p(\gamma)$) to collect all \securityRelevant{} settings $ \Gamma^{SR}_\theta = \left\{\gamma |  \gamma  \in \Gamma_{\theta} : p (\gamma )\right\}$ is a tedious and time-consuming task.
Thus, we outsource this process to organizations such as the Center for Internet Security~(CIS).
They provide a set of security-configuration guides $ \mathbb{S}_{CIS} $ (from  guides), and we use a CIS guide $ \mathcal{S}_{CIS, \theta} \in \mathbb{S}_{CIS}$ to harden our software $ \theta $.

However, there are situations in which we cannot use a guide:
First, if there is no CIS guide for a software.
Second, if there is a new update of the software and the CIS has not published its recommendations for the update yet.
Third, we have higher security requirements in our environment and need additional rules.
At Siemens, the third use case is the most important.
In all cases, the security experts need to find all \securityRelevant{} settings.
To support finding the \securityRelevant{} settings and assure that we find all \securityRelevant{} settings, we use automated classification.
False negatives, i.e., $ \gamma $ is \securityRelevant{}, but $ \neg p (\gamma )$, are more severe than false positives, because an attacker might use a non-hardened \securityRelevant{} setting to attack the system.
Classifiers should therefore avoid false negatives without labeling every setting as \securityRelevant{}.

Our running example will be the hardening of the Windows 10 OS (in the following \windowsTen ) with over 4500 settings ($ |\Gamma_{W10}| > 4500 $).
Furthermore, there is a CIS \windowsTen{} guide with over 500 rules, i.e., $ |\mathcal{S}_{CIS, W10}| \approx 500 $.
Every rule $ r $ \textit{targets} a setting $ \gamma $, which we denote with $ \varpi (r) = \gamma $ and this setting is unique, i.e., $ \varpi_{CIS, W10} $ is injective and $|\Gamma^{SR}_{CIS, W10}| \approx 500 $.
In May 2021, Microsoft released the 21H1 update for \windowsTen{} including over 300 new settings.
The security experts at Siemens now needed the new \securityRelevant{} settings, i.e., $ \Gamma^{SR}_{W10'} \setminus \Gamma^{SR}_{W10} $.

In this article, we present our solution to this problem.
We use various state-of-the-art natural language processing (NLP) to model $ p $ and classify automatically whether a setting is \securityRelevant{}.
We use the settings' descriptions in natural language as input and existing guides to identify \securityRelevant{} terms.

Our contribution is threefold.
First, we present, to our knowledge, the first approach to use NLP techniques to tackle the identification of \securityRelevant{} settings.
Second, we publish our labeled data sets\footnote{\href{https://github.com/tum-i4/Automated-Identification-of-Security-Relevant-Configuration-Settings-Using-NLP}{github/tum-i4/Automated-Identification-of-Security-Relevant-Configuration-Settings-Using-NLP}} so that other researchers can train their models on them to solve the described problem.
Third, we share the code of our models on Kaggle so that security experts can use them when they create guides.

\section{Data Set Creation}

\begin{figure}
    \centerline{\includegraphics[width=0.9\columnwidth]{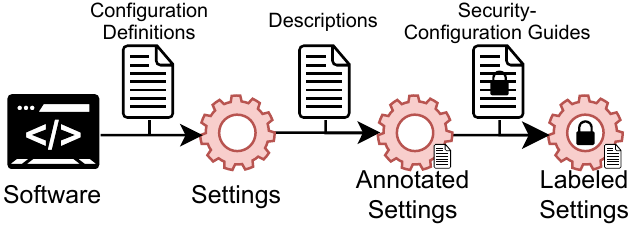}}
    \caption{Data set creation.}
    \Description{
        The graphic goes from left to right.
        Left is the software.
        An arrow leads from the software to a gear with the caption ``Settings''.
        Over the arrow, there is a document with the caption ``Configuration Definitions''.
        Another arrow leads from the settings gear to a gear with a small document attached and the caption ``Annotated Settings''.
        Over this second arrow, there is another document with the caption ``Descriptions''.
        A third arrow leads from the annotated settings to gear with a small document attached and an additional lock in the middle.
        The caption here says ``Labeled Settings''.
        Over this third arrow, there is a third document with a lock and the caption ``Security-Configuration Guides''.
    }
    \label{fig:data-set-creation}
\end{figure}

As we have only several thousand settings, we need data-efficient techniques and a labeled data set.
For a given software $ \theta $, we first needed all settings $ \Gamma_\theta $.
Second, we needed the descriptions $ \mathcal{D} $ describing their function and purpose in natural language.
Third, we needed to label each setting $ \gamma $ as \securityRelevant{} or \notSecurityRelevant{}.
One can see the three steps depicted as arrows in \autoref{fig:data-set-creation}.

As modern software can easily have thousands of settings~\cite{PEREIRA2021111044}, it is beneficial if we automate the three steps.
Therefore, we choose \windowsTen{} for our proof of concept.
In \windowsTen{}, the Administrative Templates (ATs) define most settings.
Microsoft stores these configuration definitions in so-called ADMX files, and we can automatically generate the set of settings $ \Gamma_{W10} $ out of them.
Furthermore, the \administrativeTemplates{} include all the description texts in different languages in so-called ADML files.
For our proof-of-concept implementation, we limited ourselves to English.
However, one could also investigate whether another language, e.g., Hindi, is better suited to identify \securityRelevant{} settings.
We parse the set of descriptions $ \mathcal{D} $ from the ADML files and join the definitions with the descriptions using a shared identifier to a set of settings together with their description, i.e., $ \mathcal{L} = \left\{ \left(\gamma, d \right)| \forall \gamma \in \Gamma_{W10}: \exists d \in  \mathcal{D}: id(d) = id(\gamma)  \right\}$.

To automate the third step, we need ground truth whether a setting is \securityRelevant{} or \notSecurityRelevant{}.
Here, we used CIS guides and especially the \windowsTen{} guide $ \mathcal{S}_{CIS, W10} $.
Our assumption is that a setting $ \gamma $ is \securityRelevant{} if and only if there is a rule $ r $ in the guide $ \mathcal{S}_{CIS, W10} $ that is targeting this setting $ \gamma $, i.e., $ p_{CIS, W10}(\gamma) \Longleftrightarrow \exists r \in \mathcal{S}_{CIS, W10}:  \varpi (r) = \gamma $.
We also use Siemens guides to evaluate our classifiers on guides from different organizations.
In both cases we can automatically retrieve the set of all rules $ \mathcal{S}_{CIS, W10} $ and extract for each rule the targeted setting $ \gamma $, i.e., we have a list $ \mathcal{K} = \left\{ \left( r, \gamma = \varpi ( r) \right) | \forall r \in \mathcal{S}_{CIS, W10} \right\} $.
In the end, we can construct our labeled data set by joining $ L $ and $ K$, i.e., $ \left\{ \left( \gamma, d, \left(\exists \left( r, \gamma' \right) \in \mathcal{K} : \gamma = \gamma'  \right) \right) | \left( \gamma, d, \right) \in L \right\} $
and mark for every rule $ r $ the setting $ \gamma $ as \securityRelevant{} that $ r $  targets.
The result are the labeled settings with their descriptions (see~\autoref{lst:data-set}).

The input of our implementation is the ADMX/L files of the \administrativeTemplates{} and a guide in the XML-based XCCDF format.
Microsoft regularly updates the \administrativeTemplates{}.
Thus, there are different versions of the ADMX/L files, e.g., \textit{1909} or \textit{21H2}.
We uploaded different variants into our repository.

\section{Sentiment Analysis}

We make a binary decision if a setting is \securityRelevant{} based on its text.
Therefore, our first idea was to use sentiment analysis and lexicon-based approaches in particular to solve our problem.\footnote{Code: \href{https://www.kaggle.com/tumin4/sentiment-analysis}{kaggle/tumin4/sentiment-analysis}}
Due to the descriptions' formal language, spell correction was not necessary.
First, we considered basing our classification on part-of-speech (POS) tags.
However, we found \securityRelevant{} words distributed over all POS tags.
The same holds for high frequency, as most frequent words in the \securityRelevant{} descriptions also occur frequently in \notSecurityRelevant{} ones.
We also extracted words that only occurred in \securityRelevant{} descriptions.
Several words, e.g., ``attacker", showed a relation to a security aspect, but filtering for words with a frequency of greater than five left only 12 words identifying hardly all the \securityRelevant{} settings.
As we could see subjects repeatedly mentioned in the descriptions, we used the term frequency-inverse document frequency (tf-idf) algorithm instead of the frequency.
To reduce the words to the relevant ones, we set the threshold to 0.5 and ended up with 141 words.
However, only a few came from the security domain, and so we combined the descriptions with the rationales (text explaining why one should configure a setting) of CIS rules.
In the end, we could find the 80 \securityRelevant{} words, e.g., ``microphone'' or ``trust'';
one can see all \securityRelevant{} words as word clouds in the corresponding Kaggle notebook.
Nevertheless, these words also occur frequently in the \notSecurityRelevant{} descriptions, and we constructed based on tf-idf a counterpart set of words that mark \notSecurityRelevant{} settings, e.g., ``color", but not enough to prevent a high number of false positives.
The same problem occurred when we used n-grams or named entity recognition:
The entity represents a particular case referring only to a few \securityRelevant{} settings and, therefore, contributes little to the entire classification.
Alternatively, the n-gram also appears within the \notSecurityRelevant{} descriptions and therefore could lead to \notSecurityRelevant{} descriptions being classified additionally as \securityRelevant{}.
With these findings, it becomes clear that the lexicon-based approaches lead to a large percentage of false positives, making them unsuitable in our case.
\securityRelevant{} words do not necessarily follow one after another.
Therefore, increasing the size of n-grams is not suitable as well.
Our insight here was that classifying the settings directly as \securityRelevant{} performed not as well as expected.

\begin{listing}
\begin{Verbatim}[commandchars=\\\{\}]
\PYG{p+pIndicator}{\PYGZhy{}} \PYG{n+nt}{setting}\PYG{p}{:} \PYG{l+lScalar+lScalarPlain}{Control Panel \PYGZbs{} Personalization \PYGZbs{} Force a specific background and accent color}
    \PYG{n+nt}{description}\PYG{p}{:} \PYG{l+s}{\PYGZdq{}Forces}\PYG{n+nv}{ }\PYG{l+s}{Windows}\PYG{n+nv}{ }\PYG{l+s}{to}\PYG{n+nv}{ }\PYG{l+s}{use}\PYG{n+nv}{ }\PYG{l+s}{the}\PYG{n+nv}{ }\PYG{l+s}{specified}\PYG{n+nv}{ }\PYG{l+s}{colors}\PYG{n+nv}{ }\PYG{l+s}{for}\PYG{n+nv}{ }\PYG{l+s}{the}\PYG{n+nv}{ }\PYG{l+s}{background}\PYG{n+nv}{ }\PYG{l+s}{and}\PYG{n+nv}{ }\PYG{l+s}{accent.}\PYG{n+nv}{ }\PYG{l+s}{The}\PYG{n+nv}{ }\PYG{l+s}{color}\PYG{n+nv}{ }\PYG{l+s}{values}\PYG{n+nv}{ }\PYG{l+s}{...\PYGZdq{}}
    \PYG{n+nt}{is\PYGZus{}security\PYGZus{}relevant}\PYG{p}{:} \PYG{l+lScalar+lScalarPlain}{false}
\PYG{p+pIndicator}{\PYGZhy{}} \PYG{n+nt}{setting}\PYG{p}{:} \PYG{l+lScalar+lScalarPlain}{Control Panel \PYGZbs{} Personalization \PYGZbs{} Prevent enabling lock screen camera}
    \PYG{n+nt}{description}\PYG{p}{:} \PYG{l+s}{\PYGZdq{}Disables}\PYG{n+nv}{ }\PYG{l+s}{the}\PYG{n+nv}{ }\PYG{l+s}{lock}\PYG{n+nv}{ }\PYG{l+s}{screen}\PYG{n+nv}{ }\PYG{l+s}{camera}\PYG{n+nv}{ }\PYG{l+s}{toggle}\PYG{n+nv}{ }\PYG{l+s}{switch}\PYG{n+nv}{ }\PYG{l+s}{in}\PYG{n+nv}{ }\PYG{l+s}{PC}\PYG{n+nv}{ }\PYG{l+s}{Settings}\PYG{n+nv}{ }\PYG{l+s}{and}\PYG{n+nv}{ }\PYG{l+s}{prevents}\PYG{n+nv}{ }\PYG{l+s}{a}\PYG{n+nv}{ }\PYG{l+s}{camera}\PYG{n+nv}{ }\PYG{l+s}{from}\PYG{n+nv}{ }\PYG{l+s}{being}\PYG{n+nv}{ }\PYG{l+s}{invoked}\PYG{n+nv}{ }\PYG{l+s}{on}\PYG{n+nv}{ }\PYG{l+s}{the}\PYG{n+nv}{ }\PYG{l+s}{lock}\PYG{n+nv}{ }\PYG{l+s}{screen...\PYGZdq{}}
    \PYG{n+nt}{is\PYGZus{}security\PYGZus{}relevant}\PYG{p}{:} \PYG{l+lScalar+lScalarPlain}{true}
\end{Verbatim}

\caption{Labeled settings for Windows 10, version 1909.}
\label{lst:data-set}
\end{listing}

\section{Topic Modeling}

Next, we trained a Latent-Dirichlet-Allocation (LDA) topic model to determine topics within the \securityRelevant{} descriptions.\footnote{Code: \href{https://www.kaggle.com/tumin4/topic-modeling-and-latent-dirichlet-allocation}{kaggle/tumin4/topic-modeling-and-latent-dirichlet-allocation}}
The intuition behind the LDA is that a document typically not only treats one single topic but can be rather seen as a mixture of multiple latent topics.
Once we trained the model on the \securityRelevant{} descriptions, we can calculate the probability of each description referring to a security topic.
If the probability exceeds a certain threshold, we classify the description as \securityRelevant{}.

We tokenized the descriptions, removed stop words, selected only words between 2 and 16 characters, and built the lemma and the word stem.
Of the 300 most frequent stems, we manually created a list of words that are irrelevant for the security aspect, e.g., ``kilobyte", or not specific to one topic, e.g., ``password".
We trained the LDA model on the entire collection of \securityRelevant{} descriptions.
For the evaluation, we tested the classifier on other data sets, e.g., labeled according to another guide for the same system variant or on data sets for different system versions and variants.
We built a dictionary containing all words that remained after the preprocessing for the LDA model and assigned each word an id.
The tf-idf-feature representation lists the ids with the tf-idf scores.

To optimize the performance of the model, we had to set the following parameters:
amount of topics;
passes, i.e., how often the algorithm iterates over the entire corpus to optimize the topic allocations;
$ \alpha $, i.e., a priori assumption about the document-topic distributions;
whether we would use per-word-topics;
the probability threshold.
We achieved the best results with nine topics and four passes on our training data.
For $ \alpha $, we learned an asymmetric a priori probability distribution from the description corpus and used per-word topics, i.e., we calculated a list of most likely topics for each word.
Finally, we set the probability threshold to 70\%.

The LDA topics are not as descriptive as the topics based on the CIS categories.
We would need to draw semantic relationships between the words of one CIS topic, e.g.,
typical key words relating to \textit{Data Protection} are ``send", ``collect" etc., but the words ``send" and ``collect" do not necessarily share common context words.
The knowledge about a semantic relationship between them is necessary to relate them to the same topic.
However, LDA is not capable of doing this.

For the classification, we preprocess all descriptions, transform them into the tf-idf representation, and the model returns the probabilities that the description refers to a topic.
The description in \autoref{lst:topic-prediction} has a strong relation of $\approx 80 \% $ to Topic~3.
If any topics are above our threshold, we classify the model as \securityRelevant{}.
Although the LDA obtained better results than the SA, we discovered several problems that we discuss in \autoref{sec:evaluation}.

\begin{table*}
    \caption{Classification results of the LDA-based classifier.}
    \centering
    \begin{tabular}{l l r r r r r}
    \toprule
       \multicolumn{1}{p{1cm}}{\centering \textbf{OS } \\ $\theta$} &  \multicolumn{1}{p{1cm}}{\centering \textbf{Guide \\ $ \mathcal{S}$}} &
       \multicolumn{1}{p{1cm}}{\centering \textbf{Settings \\ $ |\Gamma_\theta|$}} &
       \multicolumn{1}{p{1cm}}{\centering
       \textbf{\# of \securityRelevant{}} \\ $|\Gamma^{SR}_\theta|$} &
       \multicolumn{1}{p{1cm}}{\centering \textbf{Recall\\ (\%)}} &
       \multicolumn{1}{p{1.5cm}}{\centering \textbf{\# classified as \securityRelevant{}}} &
       \multicolumn{1}{p{1cm}}{\centering \textbf{BA \\ (\%)}} \\
       \midrule
        \windowsTen{}~\textit{1909} & CIS & 2688 & 246 & \textbf{91} & 406 & \textbf{92} \\
        \midrule
        \windowsTen{}~\textit{1803} & CIS & 2576 & 238 & 89 & 382 & 91 \\
        \windowsServer{} & CIS & 2430 & 156 & 88 & 355 & 89 \\
        \midrule
        \windowsTen{}~\textit{1909} & Siemens & 2688 & 303 &  59 & 407 & 73 \\
        \windowsServer{} & Siemens & 2430 & 192 & 80 & 355 & 85 \\
        \bottomrule
    \end{tabular}
    \label{tab:res_lda}
\end{table*}

\section{Transformer-based Machine Learning}

The promising, but not convincing results (see~\autoref{sec:evaluation}) of the topic model and the recent success of deep learning models in standard NLP tasks motivated us to train an additional model using transformer-based machine learning, namely BERT~\cite{devlin2019bert}.

For our model, we extended the sequence size to 512 tokens as the average input length in our data set was higher than the standard sequence length.
Before training the BERT model, we removed the hive duplications we recognized during our LDA experiments (see~\autoref{sec:evaluation}) from our data set.
We used a combination of over- and undersampling to deal with our imbalanced data set and the low number of \securityRelevant{} settings.
We altered all hyperparameters, e.g., batch size or k-fold, in reasonable ranges of values.\footnote{Code: \href{https://www.kaggle.com/tumin4/transformer-based-machine-learning}{kaggle/tumin4/transformer-based-machine-learning}}
We achieved the best results with a small BERT model with 2 hidden layers, a hidden size of 128, 2 attention heads, a batch size of 32, a dropout rate of 0.2, 20 epochs, and 5-fold cross-validation.
Overfitting was a noticeable problem with BERT's storing capabilities and our limited data set.
Thus, we added a dropout layer and used the AdamW optimizer with a decreasing learning rate.

\section{Evaluation \& Discussion}
\label{sec:evaluation}

\begin{listing}
\begin{Verbatim}[commandchars=\\\{\}]
description: \PYGZdq{}Windows Components. AutoPlay Policies. Turn off Autoplay.[...]\PYGZdq{}
Topic probabilities: [(0, 0.025299275), (1, 0.02649991), (2, 0.025674498), (3, 0.79593724), ..., (8, 0.026722105)]
\end{Verbatim}
\caption{Topic prediction.}
\label{lst:topic-prediction}
\end{listing}

We evaluated the performance of the LDA and the BERT model on different data sets.
The correct classification of \securityRelevant{} settings is critical, and, thus, we focus strongly on the recall.
A \textit{useful} classifier in our context should therefore score a recall close to $ 100\%$ without producing too many false positives.

Our four main research questions were:

\begin{description}
    \item[RQ1] What is the highest recall a useful LDA-based classifier trained on the \securityRelevant{} descriptions can achieve?
    \item[RQ2] Can we use our LDA-based approach as a classifier $ p $ for Siemens and CIS guides?
    Does it make sense to train publisher classifiers, i.e., $ p_{SIE}$ and $p _{CIS} $?
    \item[RQ3] Which recall/precision can our BERT-based classifier achieve on the unseen data?
    Could we achieve a sufficiently high recall of nearly 100\% to replace the manual analysis by the security experts?
    \item[RQ4] What are the main reasons for false negatives?
    \item[RQ5] What are the main reasons for false positives?
    What would we need to avoid these problems in the future?
    Can we find settings that are not part of the CIS guides but are \securityRelevant{} judged on the description?
\end{description}

\autoref{tab:res_lda} shows the classification results of the best performing LDA-based classifier on \windowsTen{} and Windows~Server~2016 (\windowsServer{}) data sets.
Our data sets are imbalanced as we have only a tiny percentage of \securityRelevant{} settings.
Therefore, we choose instead of the \textit{normal} accuracy the balanced accuracy (BA)~\cite{brodersen2010the}.
With LDA, we achieve a recall value of 91\% on the \windowsTen{}~\textit{1909} data set with a 92\% BA on our training set answering \textbf{RQ1}.
To make this clear, we used all \securityRelevant{} descriptions to identify \securityRelevant{} topics in the descriptions.
The positive result here is that we have very few false positives, i.e., the LDA-based model can differentiate between the topics of the seen \securityRelevant{} descriptions and the unseen \notSecurityRelevant{} descriptions.
One might now ask why we still missed 23 \securityRelevant{} settings and could not meet our goal of $\approx100\%$ recall.
When we investigated the 23 false negatives, we could see that the LDA-based model could not take the context of a word into account and lacked the semantical understanding necessary to classify the settings correctly.
Thus, we could train LDA-based classifiers with sufficient recall of $ \approx 100\% $, but those resulted in large numbers of false positives.
The context-sensitivity and semantical understanding motivated our BERT-based classifier as an improvement over LDA.

The LDA-based classifier has a high recall and BA values on all CIS guides ($\Delta_{recall} \leq 3 pp$, $\Delta_{BA} \leq 1 pp$), lower values on \windowsServer{}, and performs bad on Siemens~\windowsTen{}.
The results are relatively stable between different \windowsTen{} versions and \windowsTen{}/ \windowsServer{}.
Therefore, we assume that the CIS is consistent within its classification of settings based on their description.
We know that the security experts used the CIS~\windowsServer{} as a basis for the Siemens~\windowsServer{} guide explaining the relatively good performance.
After seeing the bad results on Siemens~\windowsTen{}, we investigated the difference between the CIS and the Siemens guide.
We found many settings targeted only in one guide but not in the other;
even if such a setting was in the training data, the classifier could not predict it correctly.
With this in mind, training a global $ p $ does not make sense, but a publisher classifier $ p_{CIS} $ is useful.
With the limitation to two publishers and Windows-based OSs, we could answer \textbf{RQ2}.

\autoref{tab:res-bert-vs-random} shows the result of our BERT-based classifier.
As we present the first automated classification approach, we compare it with the best-performing dummy classifier, i.e., randomly classifying $x\%$ of the settings as \securityRelevant{}, as a baseline.
Although the dummy classifier has a better recall, its precision is only 11\%, thus producing too many false positives.
In precision and F1, the BERT classifier outperforms the baseline by 30pp respectively 24pp.
However, our classifier misses more than half of the \securityRelevant{} settings in the test data.
\autoref{tab:res-bert} shows how our classifier performed on our other data sets.
As the data sets share settings, we made sure that we used in the test data set no settings that we previously used in training.
Nevertheless, although trained on CIS~\windowsTen{}~\textit{1803}, our classifier performs best on the \windowsTen{}~\textit{1909} with a 60\% recall and 46\% precision.
Our explanation for the good result on the newer version is that CIS marks some new settings as \securityRelevant{} and changes some old settings from \notSecurityRelevant{} to \securityRelevant{}.
However, CIS's updates to their guides make them more consistent, at least to what the classifier has learned from the descriptions.
As we want to use the classifier in this use case of a new software version, we see this number as a basis for the future, but in the end, we are still far away from 100\% recall.
Therefore, we cannot replace the manual analysis of security experts, and we could not fulfill the second part of \textbf{RQ3}.

Going through the false negatives of our classifiers, we identified four main classification problems.
Unique settings, short descriptions, descriptions with a vocabulary spread over multiple topics, and linked settings.
An example of the first group is the setting \textit{Enable Windows NTP Server}.
The targeting rule's rationale state that it is \securityRelevant{} for the validity of timestamps used, e.g., in authentication procedures.
However, the setting's description neither includes ``clock" nor ``synchronization" and neither the LDA nor the BERT-based models label it as \securityRelevant{}.
An example of the second group is \textit{Allow Cloud Search}.
Here, the description only consists of one sentence, and we cannot assess the topic.
The third group is settings whose description is \securityRelevant{} according to two or more topics.
However, no single probability is over the threshold.
Our LDA classifier assigns the setting \textit{Allow user control over installs}  to 51\% to Topic~3 and 35\% to Topic~4.
Thus, we classify it wrongly as not \securityRelevant{}.
The fourth group is settings that often occur in other settings' descriptions.
Several \notSecurityRelevant{} settings mention the \securityRelevant{} setting \textit{Prevent enabling lock screen slide show}.
Thus, the classifier deducts that this setting is \notSecurityRelevant{}.
Linked settings also cause false positives if multiple \securityRelevant{} settings mention a \notSecurityRelevant{} setting.
The four presented groups answer \textbf{RQ4}.

Next, we went through the classifiers' false positives.
We could identify four groups of common problems:
Overruled settings, hive duplication, correction candidates, and context-specific meanings.
The first group is settings with \securityRelevant{} descriptions.
Nevertheless, they become ineffective if another setting is enabled or disabled.
The setting \textit{MS Support Diagnostic Tool \textbackslash Configure execution level} states that it takes no effect if the ``scenario execution policy" is configured.
We would need a semantic model of the settings' relations to avoid such false positives.
The second group is settings existing both in the Computer \textbf{and} the User hive.
They usually have the same description, but the Computer setting has precedence over the User setting.
Thus, the CIS marks the Computer setting as \securityRelevant{} and the User as \notSecurityRelevant{}.
However, there are settings like \textit{Always install with elevated privileges} stating that we should enable this policy in both hives.
Thus, we needed to know which settings are essential on both hives to prevent these false positives.
Since we trained the BERT-based model after the LDA evaluation, we removed this problem there.
The third group is settings that indeed seem \securityRelevant{}, e.g., because we found similar written \securityRelevant{} descriptions.
One example here is the \textit{Prohibit non-administrators from applying vendor signed updates} setting.
We do not know whether the CIS overlooked this setting or deliberately chose to omit this setting, e.g. because the impact is meager.
The fourth group is settings that have words that are only in some contexts \securityRelevant{}, e.g., \textit{Prevent Application Sharing in true color}.
``Application" and ``Sharing" appear in many \securityRelevant{} descriptions, but here, this color setting is \notSecurityRelevant{}.
To filter out those rules, we would need to take the context of the words more into account.
Only the third group provides candidates for the new rule.
However, as we do not know whether the CIS forgot them or omitted them, we cannot answer the second part of \textbf{RQ5}.

\begin{table}
    \caption{Performance of the BERT and the dummy classifier on CIS Windows 10, version 1803.}
    \centering
    \begin{tabular}{ l r r r }
        \toprule
       \multicolumn{1}{p{1cm}}{\centering \textbf{Classifier} } &
       \multicolumn{1}{p{0.7cm}}{\centering \textbf{Recall}} &
       \multicolumn{1}{p{1cm}}{\centering \textbf{Precision}} &
       \multicolumn{1}{p{0.5cm}}{\centering \textbf{F1}}
       \\
       \midrule
        BERT &   0.44 & \textbf{0.41} & \textbf{0.42} \\
        Uniform & \textbf{0.54} &   0.11 &  0.18  \\
        \bottomrule
    \end{tabular}
    \label{tab:res-bert-vs-random}
\end{table}

Our evaluation shows that our classifiers could detect many settings correctly, but not enough for our use case.
The main problem with the descriptions is that they should inform a user about the setting not a security expert about the setting's security implications.
Our findings suggest that NLP techniques like the LDA topic model alone cannot replace the security experts and their domain knowledge in this task.

\section{Related Work}

Research about configuration is an essential part of the software engineering~\cite{bhagwan2021learning,nguyen2021gentree} as well as the security domain~\cite{dietrich2018investigating,stoeckle2022hardening}.
Stöckle et al. demonstrated how one could use NLP to implement guides efficiently~\cite{stoeckle2020automated}.
Most relevant for the problem of identifying \securityRelevant{} settings is sentiment analysis, where we classify documents as being positive or negative, depending on the expressed sentiment~\cite{hemmatian2019a}.
We limited ourselves to SA approaches that do not need much data.
Qiu et al. start from a seed lexicon containing a few meaningful features and expand it via the exploitation of a specific characteristic~\cite{qiu2011opinion}.
They use dependency rules to extract features from the data set and add words iteratively to the seed lexicon that occur in a particular dependency relation to a word from the seed lexicon.
The lexicon-based approaches build on the assumption that specific words express either one of the opposing sentiments, i.e., \textit{good} is characteristic for the positive and not for the negative sentiment.
However, our evaluation shows that the assumption does not hold for the vocabulary of settings' descriptions.

\section{Conclusion}

We constructed labeled data sets for security-relevant configuration settings.
We motivated our decision to train an LDA topic model and a BERT-based model to classify \securityRelevant{} settings.
Our evaluation could achieve good results on the different data sets.
The required recall of close to 100 \% due to the security implications could not be met.
Therefore, our approach cannot replace security experts going through the settings.
Nevertheless, it can provide good support for them.
We published our labeled data sets so that other researchers can use them for training better models in the future.

Based on our results, we propose several improvements for the configuration hardening:
First, we need data sets with settings, descriptions, and security relevancy for more systems, e.g., Linux-based systems or applications.
Second, software vendors should improve the settings' descriptions and add security implications.
Third, it would be better if the software vendors tag all \securityRelevant{} settings directly in a machine-readable way, e.g., in the ADMX, so that we would not need NLP techniques to extract it from the natural language texts.
Fourth, the software vendors could provide machine-readable security-configuration guides, e.g., in XCCDF or Scapolite, along with their software.
With these guides, security-aware users could harden their systems directly during the installation and make them secure from day one.

\begin{table}
    \caption{Classification results of the BERT-based classifier.}
    \centering
    \begin{tabular}{ l l r r r }
    \toprule
       \multicolumn{1}{p{1cm}}{\centering \boldmath$\theta$} &
       \multicolumn{1}{p{1cm}}{\centering
       \boldmath$\mathcal{S}$} &
       \multicolumn{1}{p{0.7cm}}{\centering \textbf{Recall}} &
       \multicolumn{1}{p{1cm}}{\centering \textbf{Precision}} &
       \multicolumn{1}{p{1cm}}{\centering \textbf{F1}}
       \\
       \midrule
        \windowsTen{}~\textit{1803} & CIS & 0.44 & 0.41 & 0.42  \\
        \windowsTen{}~\textit{1909} & CIS & \textbf{0.60} & \textbf{0.46} & \textbf{0.52} \\
        \windowsServer{} & CIS & 0.49 & 0.28  & 0.35 \\
        \windowsTen{}~\textit{1909} & Siemens & 0.48 & 0.33 & 0.39  \\
        \windowsServer{} & Siemens & 0.48 & 0.43 & 0.45  \\
        \bottomrule
    \end{tabular}
    \label{tab:res-bert}
\end{table}

\bibliographystyle{ACM-Reference-Format}
\bibliography{main}

\end{document}